\documentclass[]{spie}  %>>> use for US letter paper
\usepackage[]{graphicx}
\usepackage{amsmath, amsfonts}
\usepackage{mcite}
\usepackage{dsfont}

\title{A geometric view of quantum cellular automata  } 

\author{Jonathan R. McDonald\supit{a}, Paul M. Alsing\supit{a}, and Howard A. Blair\supit{b} 
\skiplinehalf
\supit{a}Air Force Research Laboratory, Information Directorate, Rome, New York 13441; \\
\supit{b}Department of Computer Science and Electrical Engineering, Syracuse University, Syracuse, New York
}

\authorinfo{Further author information: (Send correspondence to
  J.R.M.)\\J.R.M.: jmcdnld@gmail.com\\  P.M.A.: paul.alsing@rl.af.mil\\  H.A.B.: blair@ecs.syr.edu }
\newtheorem{mydef}{Definition}
\newcommand{\J}{\ensuremath{{\cal J}}}

\begin{document} 
  \maketitle 

%%%%%%%%%%%%%%%%%%%%%%%%%%%%%%%%%%%%%%%%%%%%%%%%%%%%%%%%%%%%% 
\begin{abstract}
  Nielsen, {\em et al.} \cite{Nielsen:QC2006, Nielsen:QC2008}
  proposed a view of quantum computation where determining optimal
  algorithms is equivalent to extremizing a geodesic length or cost
  functional.  This view of optimization is highly suggestive of an
  action principle of the space of $N$-qubits interacting via local
  operations.  The cost or action functional is given by the cost of
  evolution operators on local qubit operations leading to causal
  dynamics, as in Blute {\em et. al.}\cite{Blute:2003}  Here we propose a view of
  information geometry for quantum algorithms where the inherent
  causal structure determines topology and information
  distances\cite{Zurek:1989dist, Schumacher:1991dist} set the local
  geometry.  This naturally leads to geometric characterization of
  hypersurfaces in a quantum cellular automaton.  While in standard
  quantum circuit representations the connections between individual
  qubits, i.e. the topology, for hypersurfaces will be dynamic,
  quantum cellular automata have readily identifiable static
  hypersurface topologies determined via the quantum update rules.  We
  demonstrate construction of quantum cellular automata geometry and
  discuss the utility of this approach for tracking entanglement and
  algorithm optimization.
\end{abstract}

%>>>> Include a list of keywords after the abstract 


%%%%%%%%%%%%%%%%%%%%%%%%%%%%%%%%%%%%%%%%%%%%%%%%%%%%%%%%%%%%%
\section{Introduction}
\label{sec:intro} 

When Richard Feynman \cite{Feynman:QC1982} introduced the conceptual
foundations of quantum computation it was an attempt to ask the
question, ``What dynamical rules must a computational device admit to
efficiently--and {\em exactly}--simulate Nature?''  Feynman only
requires physical processing of information through {\em local}
interaction, {\em one} of the primary features of cellular automata.
In this manuscript we follow this lead and want to look at a local
geometric view of computation that represents the causal and
information dynamics involved in the computation.  While we develop a
generic approach for quantum computation, we will apply it directly to
cellular automata.  Cellular automata are particularly interesting
given their inherently local behavior and the homogeneity in space and
time.  As a result the connectedness of the qubits is static and
defines a clear notion of how one qubit can effect any other qubit in
the system. 

A goal of computation is to take a system of qubits in some initial
configuration and output a configuration from which we can ascertain
the answer to some question of interest.  Feynman's goal was to have a computation
that could effectively simulate a quantum system, but we can ask more
general questions.  In quantum computation, this reduces to accurately
and efficiently simulating the action of some arbitrary, but known,
unitary operator.  One can then ask, given a unitary operator acting
on $N$ qubits, what is the optimal decomposition into local $1$, $2$,
or higher-order qubit interactions?  One approach to this was
introduced by Nielsen {\em et. al.}\cite{Nielsen:QC2006,
  Nielsen:QC2008} using the language of differential geometry in
which the optimal algorithm is obtained via the minimal
geodesic--according to the inherent Riemannian metric--to the desired
unitary.  While this approach is reminiscent of an action principle
which can be extremized and equations of motion extracted, it is
somewhat at odds with typical physical models with actions which are
usually aimed at describing some trajectory through configuration
space of the physical system, i.e. the qubits.  Indeed the geodesics
in the space of operators are dual to trajectories of the qubits in
configuration space.  It is this latter trajectory we hold as a focus
for determining the efficient simulation of a given unitary operator.

The beauty of the minimal geodesic interpretation is in its geometric
in origin.  However, the geodesic approach's natural arena is the
space of operators.  It is reasonable to think, then, that there
exists a dual approach on the configuration of qubits in the
computation.  In this manuscript we developed a view of computation as
an evolution of a geometry of qubits and examine the case of QCA. In
Section~\ref{sec:found} we will present the foundations relevant to
construction of surface geometries of qubits in a computation,
including topology construction and metric information.  Then in
Section~\ref{sec:qca} we examine QCA as a initial test and toy model
of this approach in 1D QCA.  We end with a discussion of conclusions
from the QCA simulations and future directions.

%%%%%%%%%%%%%%%%%%%%%%%%%%%%%%%%%%%
%%%%%%%%%%%%%%%%%%%%%%%%%%%%%%%%%%%
\section{Foundations: What is the geometry of a system of qubits?}
\label{sec:found}

Given a computation that specifies a set of qubits and local
interactions.  We propose that the content of the computational
structure allows for the construction of a low-dimensional manifold
(${\rm dim} ({\cal M}) \ll 2^{N}$) for the qubits at any given moment
of simultaneity, spatial slice, in the computation.  Moreover, the the
topological properties of such a manifold represents the
connectivity/causal properties of the computation in the future of the
given slice.  Geometric properties on the topology may be assigned in
a number of ways, but a reasonable requirement seems to be that
distances should correspond to the informational correlation of local
states in the computation, independent of that actual distribution in
Hilbert space. Once we have prescription for the computational
geometric state, we can construct the geometric evolution of the
computation from foliations of the evolution.

We examine in this section the key ingredients to computational
manifold construction. A topology tells us which qubits are localized
near other qubits in the computation.  This is a property of the
computation itself encoded in the interconnectivity of the local
quantum operations.  We interpret each qubit as a point in our
topology. What is the topology on the set of qubits?  It is true that
there is the natural topology on the composite Hilbert space, but such
a topology is not necessarily characteristic of the computation and
does not provide any simplification over the $2^{N}$-dimensional
Hilbert space.  Instead the topology of our computational geometry is
to be derived from the causal history of qubits in the computation.

Given a topology, the geometric content is contained in the metric
information of the (pseudo)-Riemannian manifold.  In the computational
manifold, metric information corresponds to local distances between
neighboring neighborhoods of qubits (or between ``nearby'' qubits).
Such a distance represents the degree to which information contained
in two distinct neighborhoods is correlated.  Such a measure based on
correlated information allows one to use geometric content as a probe of
classical and quantum correlations.\cite{Schumacher:1991dist} Below we
review the information distance on quantum states and describe a
methodology for topology construction on which to assign geometric
content.

\subsection{Information Distance}
\label{subsec:dist}

\begin{figure}[h]
   \begin{center}
   \begin{tabular}{c}
   \includegraphics[height=1.5in]{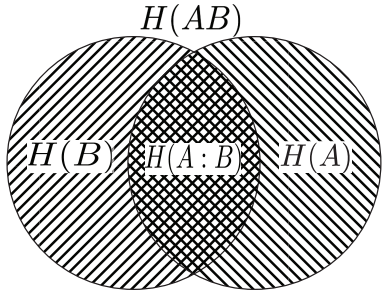}
   \end{tabular}
   \end{center}
   \caption[example] 
   { \label{fig:InfoVenn} The relationship between
     marginal, joint and mutual information.}
   \end{figure}

We have already outlined the broad proposal being investigated, now we
must generate the necessary mathematical framework to assign a
manifold to an instance in the computation.  Given the topology, the
closeness of two neighboring qubits is determined by the metric
information.  Two neighboring qubits neighbor one another if
information flow from one qubit to another need not flow through
intermediary qubits.  However, the topological relation gives no
content on the relative distance between any two such topologically
nearby qubits.  One way to assign quantitative geometric meaning to
two information carrying systems is through correlation information.
Two qubits can be said to be close if they are strongly correlated and
far if they are uncorrelated.  One  measure of the correlation
between two information carrying systems is Zurek's information
metric\cite{Zurek:1989dist}
\begin{align}\label{eq:InfoDist}
\begin{split}
\delta(A, B) &= H(A|B)+H(B|A)  \\
&= H(A B) - H(A: B)\\
&= 2H(A B) - H(A) - H(B) 
\end{split}
\end{align}
where $H(A)$ is the Shannon entropy on system $A$, $H(AB)$ is the
joint information, $H(A|B)$ is the conditional information, and $H(A:
B) = H(A)+H(B)-H(AB)$ is the mutual information.  On classical
systems, this distance satisfies all the criteria for a distance
measure to be a metric. The utility of this measure is that it
utilizes the joint and local states to provide a true measure of the
correlated content of the two subsystems.  We know of no other
distance measure that identifies the correlation content between two
subsystems of a composite system in a known state.  The Zurek
information distance is the measure that gives a direct indication of
the information overlap between $A$ and $B$, given the joint state of
$AB$.  This is evident from the second line of (\ref{eq:InfoDist})
which captures the essence of information distance (see
Figure~\ref{fig:InfoVenn});
\begin{equation} \label{eq:ConceptInfoDist} 
\left( \begin{array}{c}
      \text{Information}\\ \text{Distance}\end{array}\right) =
  \left( \begin{array}{c} \text{Information content of}\\ \text{the
        joint system, AB}\end{array}\right) - \left( \begin{array}{c}
      \text{Information common}\\ \text{to both A \&
        B}\end{array}\right)
\end{equation}

   Extension from the classical information metric with Shannon
   entropy to the quantum, using von Neumann entropy, is not
   automatic.  In the above, the information distance relies the
   mutual information or the conditional information in two of the
   classically equivalent formulations.  Since conditional information
   on quantum states presumes some basis of measurement variation of
   which may generate distinct conditional information outcomes, there
   must be some care in the information distance as applied to quantum
   states.  The discord,\cite{Zurek:Discord} Zurek's measure of the
   quantumness of a state, is typically defined as the difference
   between classically equivalent measures of mutual
   information. However, the essence of these differences between two
   such measures applied to quantum states is in the violation of
   Bayes' theorem, $S(A|B) \neq S(AB)-S(B)$, and the quantum disparity
   between classically equivalent measures of conditional information.
   Therefore, if we focus on the conceptual meaning of the information
   distance as laid out in (\ref{eq:ConceptInfoDist}), then we can
   avoid any problems arising from quantum discord.  If we
   take as the fundamental measure $\delta(A, B) = 2S(AB) - S(A) -
   S(B)$, then we avoid such problems completely.

\begin{figure}[h]
   \begin{center}
   \begin{tabular}{c c c}
   \includegraphics[height=1.5in]{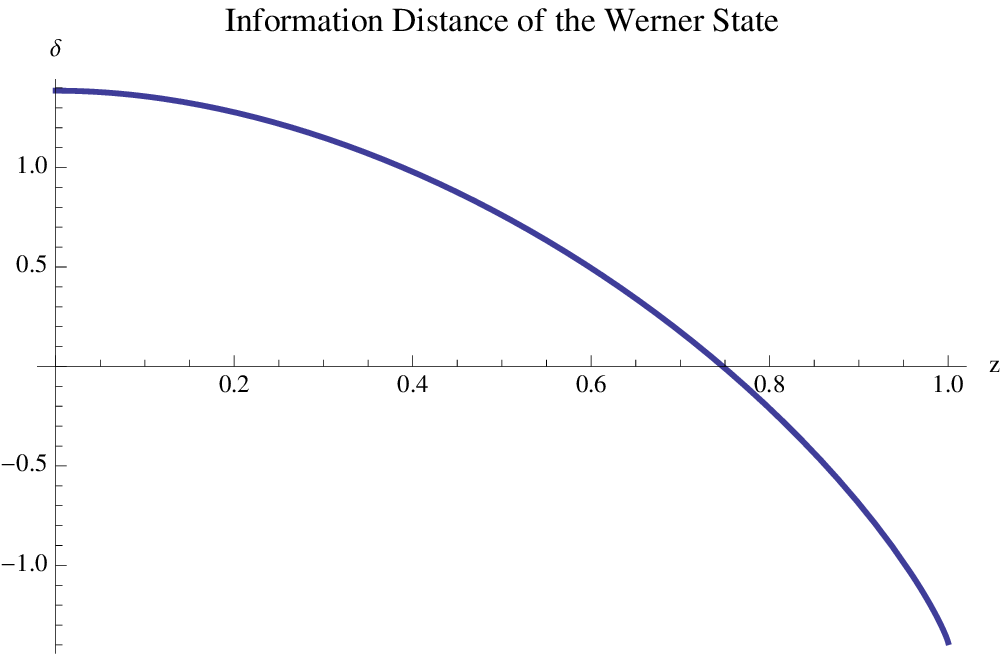} &\hspace{.5in}
   & \includegraphics[height=1.5in]{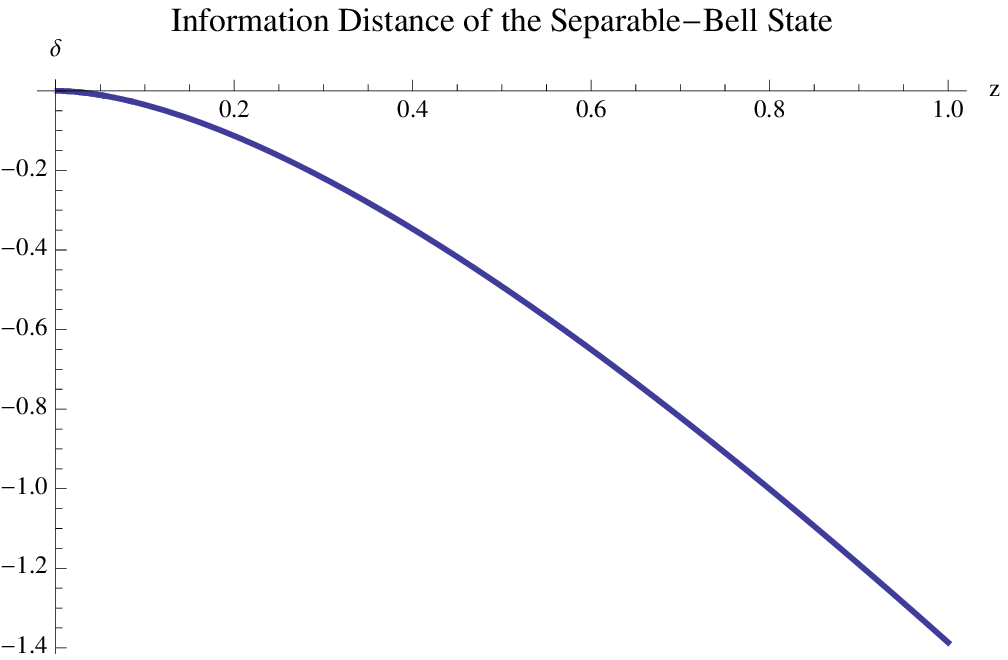} \\
(a) & & (b)
   \end{tabular}
   \end{center}
   \caption[example] 
   { \label{fig:WernerInfoDist} Information distance with classical
     and quantum correlations: (a) Here we show the mixture of the
     maximally mixed state with the Bell state $|\beta_{+}\rangle =
     (\sqrt{2})^{-1}\left(\left|00\right\rangle +
       \left|11\right\rangle\right)$.  The $z=0$ axis corresponds to
     the maximally mixed state while $z=1$ corresponds to the pure
     Bell state.  As we can see, the classical correlations quickly
     overwhelm nonlocal quantum correlations in the information
     distance. (b) Here we plot the information distance for the state
     $|\psi\rangle = \sqrt{1-z}|00\rangle
     +\sqrt{z}\left(|01\rangle+|10\rangle\right)$.  Only for the
     separable state is the information distance $0$, as can be checked with
     the Peres-Horodecki criterion.  For the state with nonlocal
     quantum correlations, the distance is always negative. }
   \end{figure}

   Another peculiarity, though essential for our purposes, arises with
   the use of the $\delta$ information distance.  When applied to pure
   quantum states, the information distance yields $0$ for any
   separable state and a negative distance in the presence of any
   non-local correlations (Figure~\ref{fig:WernerInfoDist}).  When
   applied to measurement outcomes on singlet states, it is known that
   this measure fails to satisfy the necessary properties of a
   (Riemannian) metric.\cite{Schumacher:1991dist} On general quantum
   states (without conditioning on measurements), the distance
   satisfies all properties of the metric except in relation to
   positive-definiteness and the triangle inequalities.  However, if
   we instead think of the information distance as a measure on a
   pseudo-Riemannian space we can recover the notion of a metric.
   First, we consider the case of pure, separable states as genuine
   zero distance.  This is reasonable since there exists some basis
   (unique to each party) such that the probability distribution of
   measurement outcomes will be identical for the two systems.  On the
   space of 2-qubit density matrices there still exists a distinct set
   of null distances.  It is easily seen from the definition that a
   composite system will generate null distance between two of its
   constituent parts when $S(AB) = S(A) = S(B)$.  Hence, we have a
   break down into positive distances,
   $S(AB)>\frac{1}{2}\left[S(A)+S(B)\right]$, negative distances,
   $S(AB) < \frac{1}{2}\left[S(A) +S(B)\right]$, and null distances.
   Figure~\ref{fig:WernerInfoDist} shows the information distance as
   applied to the Werner state. We see the explicit crossing of the
   null surface long before the system becomes separable but as the
   classical uncertainty in the joint system play a more important
   role than the quantum, non-local correlations.  It is not clear at
   this point the meaning of such a situation; however, there are
   indications that one may be able to distinguish between the
   classical and quantum correlations through the mutual
   information.\cite{Vedral:1997} These issues aside, the information
   distance assigns a metric on a topological scaffolding for our
   computation.  We need only be wary of assigning a particular
   importance to positive distances. However, negative information
   distances provide a clear marker for non-local correlations in the
   system.

\subsection{Causal Foliations of Computations}
\label{subsec:causal}

\begin{figure}[h]
   \begin{center}
   \begin{tabular}{c c c }
     \includegraphics[height=1.75in]{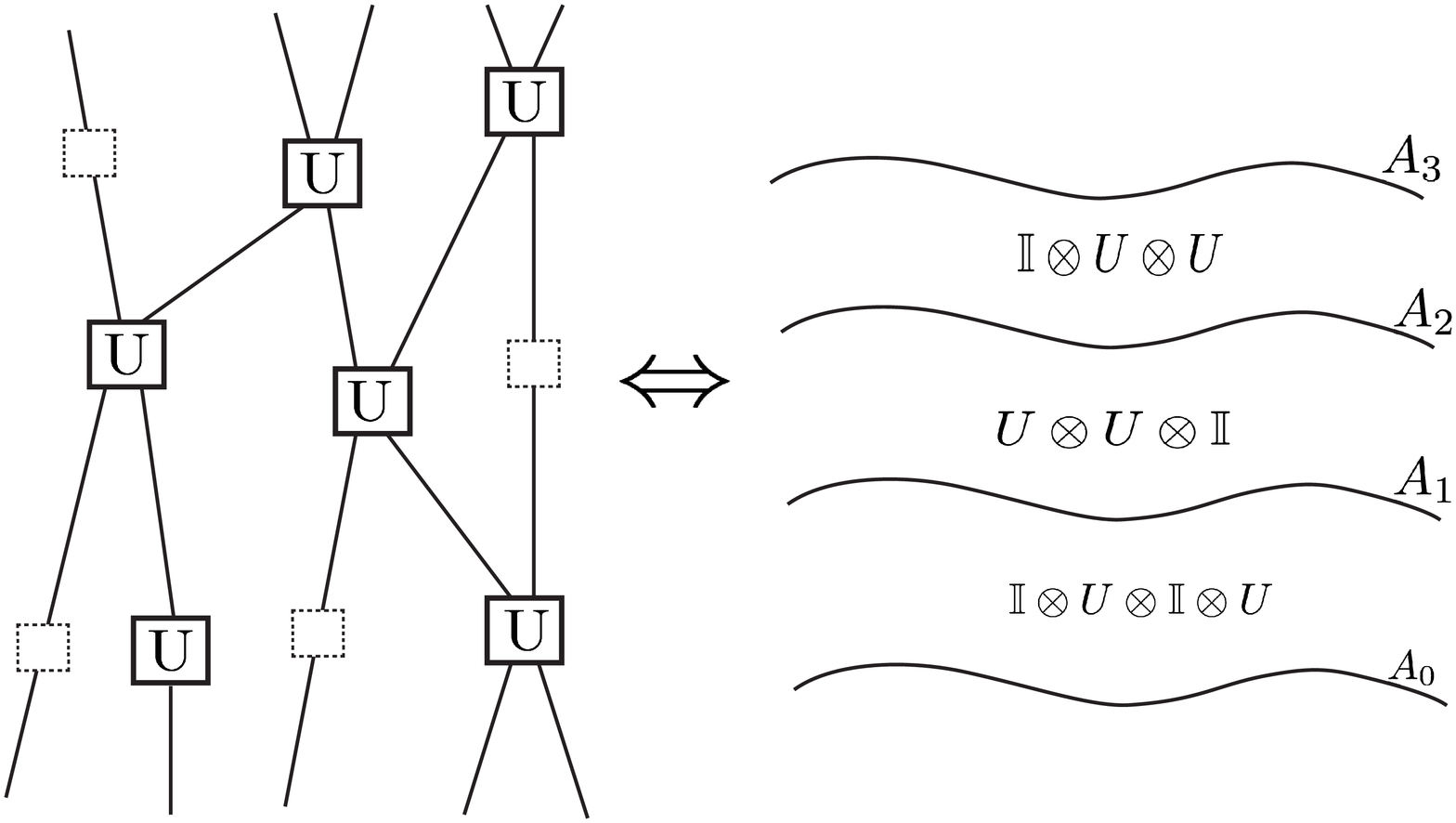} & \hspace{.5in} &
     \includegraphics[height=1.75in]{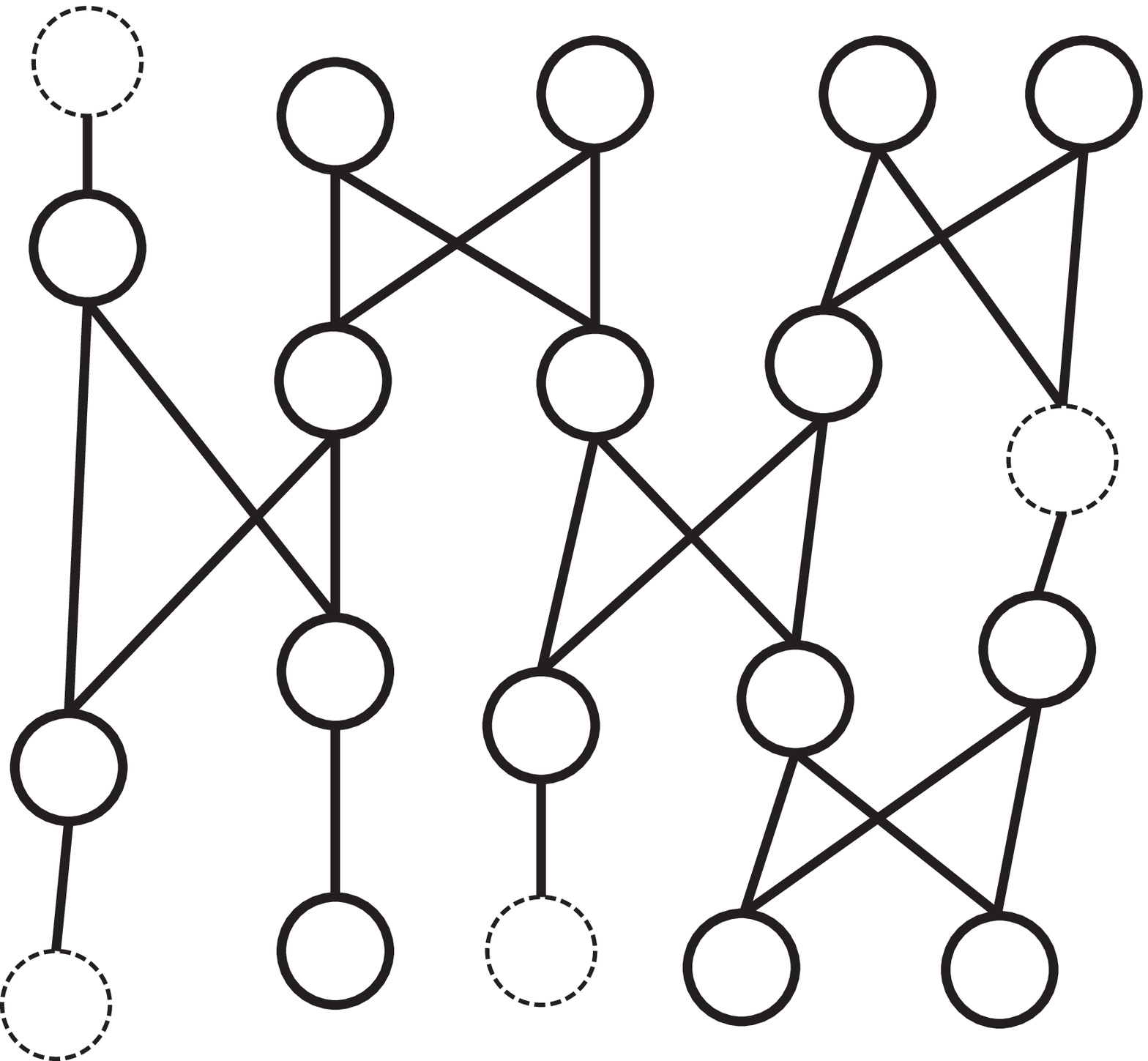}\\
     (a)& & (b)
   \end{tabular}
   \end{center}
   \caption[example] 
   { \label{fig:Foliation} (a) The computation induces an inherent
     partial order on the local unitaries (boxes) used to decompose a
     given global operator.  Hilbert spaces (wires) connected by two
     unitaries indicate the partial ordering relation. This partial
     order induces foliations on the computation.  Identity operators
     (dashed boxes) can be introduced as needed in order to construct
     non-degenerate surfaces in the foliation.  (b) The partial order
     on the unitaries induces a partial order (lines) on the Hilbert
     spaces (circles).  Whenever an identity operator is introduced, a
     copy of the Hilbert space (dashed circle) is made and an ordering
     relation is introduced.  }
   \end{figure}

We now shift our focus to construction of the scaffolding on which the
above information metric is to be applied. We first review how one
constructs foliations, a one-parameter family of non-causally
correlated surfaces, of a quantum computation.  For any quantum
computation there exists a partial order (or causal order) on the
unitary operators defining the dynamics given by the ordering of the
application of the operators to the qubits.  A partially ordered set
(poset) is a set of elements endowed with an order $x\preceq y$,
denoting $x$ precedes $y$, with the following properties;
\begin{subequations}
\begin{equation}
x\preceq x
\end{equation}
\begin{equation}
x\preceq y \  \& \  y\preceq x \Rightarrow x =  y
\end{equation}
\begin{equation}
x\preceq y\  \& \  y\preceq z \Rightarrow x \preceq z.
\end{equation}
\end{subequations}
For two unitaries $U_{i}$ and $U_{j}$ in a quantum dynamical system,
we can say that $U_{i} \preceq U_{j}$ if the commutator $\left[U_{i},
  U_{j}\right]$ is nonzero and $U_{i}$ causally precedes $U_{j}$. This
partial order on operators imposes a dual partial order on the Hilbert
spaces between successive unitary operators\cite{Blute:2003}
(Figure~\ref{fig:Foliation}).  From here  will focus on the poset
of Hilbert spaces.  The poset on unitaries will play a central role
shortly when we construct topologies on slices of a foliation.

   The poset encodes the dynamical history of the computation. The
   content of the poset generates enough information to determine a
   foliation, and all foliations, of the computation.  Similar to
   general relativity with its natural causal order, there will, in
   general, be many foliations of the poset. For the purposes here we
   need only consider one such foliation.  Different foliations of the
   same computation are equivalent under suitable local evolution.
   We first require a notion of a single surface in the foliation.

\begin{mydef}
  An anti-chain, $A$, in the poset is a subset such that no two
  elements are related through the partial order; 
$$ A = \{ x, y | x\neq y \Rightarrow\text{ neither } x\preceq y \text{ nor } y\preceq x\}.$$ 
A maximal anti-chain, $\tilde{A}$, is an anti-chain such that all
elements not in $\tilde{A}$ are related through the partial order to
at least one element in $\tilde{A}$.\\\\
\end{mydef}
A maximal anti-chain of Hilbert spaces forms a global Hilbert space of
the computation characterizing a given state of the computation.  A
computation can then be decomposed into a set of maximal anti-chains:
\begin{mydef}
  A foliation on a poset is a collection of maximal anti-chains ${\cal
    F} = \{\tilde{A}_{i}\}$
  such that  $\tilde{A}_{i} \cap \tilde{A}_j = \emptyset$ for all 
  $\tilde{A}_i, \tilde{A}_j \in {\cal F}$. \\\\
\end{mydef}
Blute {\em et. al.}\cite{Blute:2003} define the process of forming
foliations using {\em locative} slices, which will not generally be
maximal anti-chains.  We follow a similar procedure in which maximal
anti-chains play the pivotal role and may be thought of as complete
spatial hypersurfaces/slices of the computation.  These surfaces form
the global, computational Hilbert space and a state assigned to such a
surface is the global state of the $N$ qubits.  Since we only examine
maximal anti-chains we will implicitly assume $A_{i} = \tilde{A}_i$
for notational simplicity.  Assume a maximal anti-chain, $A_{0}$, on
the Hilbert spaces, which we will refer to as the initial value data.
Generally this set will be taken to be the minimal set with respect to
the partial order.  $A_{0}$ is the input to a set of unitary operators
in the computation. Whenever necessary one can split a Hilbert space,
${\cal H}_{i}$, into two Hilbert spaces, ${\cal H}_{i_1}$ and ${\cal
  H}_{i_2}$ such that ${\cal H}_{i_1} \preceq {\cal H}_{i_2}$, with
the identity as the operator relating the two spaces.  In this case,
every state in a Hilbert space on slice $i$ is causally determined by
unitary operators acting on states from slice $i-1$.  It is clear that
the need for the introduction of identity operators arises from the
definition of a foliation in which no two slices can overlap.  Without
introduction of these trivial mappings, two such slices may be
degenerate on a given qubit.

From the global system, we obtain local density matrices through the
partial trace;
\begin{equation}\label{eq:Ptrace}
\rho_{i_1 \cdots i_n} = \text{Tr}^{j_1 \cdots j_m} \rho_{i_1 \cdots i_n j_1 \cdots j_m}.
\end{equation} 
Since we have implicitly assumed only unitary operators are those
allowed in construction of the foliations, we have not needed to
distinguish the case where the outcomes of measurements may influence
the state.  However, it is worth noting that for quantum dynamics
where measurements are necessary, one must take care with
(\ref{eq:Ptrace}).  In particular, it is necessary to ensure that
knowledge of outcomes for a measurement that is not causally ordered
with respect to a Hilbert space cannot affect a state in that Hilbert
space.\cite{Blute:2003} For the present work with unitary operators,
this is not a concern and the above reduction via partial trace is
sufficient.

\subsection{From Foliations to Dynamical Geometries of Qubits}
\label{subsec:fol2geom}

Defining foliation information is insufficient content for assigning
geometric information of the kind we seek.  Instead, we have to
construct the scaffolding onto which distance information will rest.
To each slice of a foliation we must be able to assign a corresponding
topology.  It is known that complete causal information can determine
the topology of a causal manifold\cite{Penrose:1967} and this has been
proposed as a way to measure topology of space-like hypersurfaces in
causal set theory.\cite{Major:2007} The tools used here are akin to an
expanding field of computational homology\cite{Mischaikow:CH} and
persistent homology.\cite{Zomorodian:CompTopo09, Carlsson:2009} We
will outline some of the basic ideas and apply them directly to the
computational foliation.

We follow Major {\em et. al.}\cite{Major:2007} in our initial
construction of the topology for an anti-chain.  Implicit in this
scheme we will make the assumption that local operations act on
finitely many qubits in the system.  This is natural for our
construction since it implies that a given local state a qubit only
depends on a finite number of interactions with finitely many qubits in
the past.  We will then modify the approach as necessary to obtain an
appropriate simplicial complex on which to assign a metric. Start with
a maximal anti-chain, $A_{j}$, on which we desire to prescribe a
topology. We first make remarks on notation. We denote the set of
points causally influenced by $x$ as $\J^{+}(x) = \{ y | y\preceq x\}$
. The set of points causally influencing $x$ is given by $\J^{-}(x) =
\{ y| x\preceq y\}$. Because we have included reflexitivity, $x\preceq
x$, in our partial order $\J^{+}(x)\cap \J^{-}(x) = \{x\}$. A
thickened anti-chain $A^{i}_{j}$ of thickness $i$ is the set
\begin{equation}  \label{eq:thickAC}
A_{j}^{i} = \{ p | \ \ \text{card}\left(\J^{-}(p) \cap \J^{+}(A_{j})\right) \leq i+1 \in
  \mathbb{N} \}
\end{equation}
Our assumption earlier that the local unitaries in our computation
acts only on finitely many qubits is necessary to ensure that we
obtain thickened anti-chains for $0<i<\infty$.  Moreover, the set $A_i^j$ forms
a subposet of the global poset.  Assume we take a given $i <\infty$;
there is a set of maximal elements, $M_{j}^{i} \subset A_{j}^{i}$,
since the thickened anti-chain is of finite thickness, i.e. there are
finitely many points in the future of any point in $A_{j}$.  For any
$m_{k}\in M_{j}^{i}$, we define the set $P(m_{k}, A_j):= P_k=
\J^{-}(m_{k}) \cap \J^{+}(A_{j})$.  The shadow, or projection, of the
causal past of $m_{k}$ onto $A_{j}$ is given by $\bar{A}_{k} = P_{k}
\cap A_{j}$.  A simplicial complex on $A_{j}$ is generated by the
following rules: (1) each $\bar{A}_{k}$ is assigned a vertex $v_{k}$
and (2) any non-empty union of $(n+1)$ of the $\bar{A}_{k}$ is
assigned a $n$-simplex.  The homology can then be computed on each
of these simplicial complexes.  Each thickened anti-chain generates a
simplicial complex.  We look for stable topological properties as
identifiers of the topology of $A_{j}$ over the set of simplicial
complexes generated by increasing the thickness of the anti-chain.
For a connected computation, i.e. one where information can transfer
information from one segment of the computation to any other, we first
allow the manifold to become connected prior to examining stable
homology groups.  This may seem arbitrary; however, if the computation
never achieves a single connected component we are justified in
treating the system as two distinct and separable computations from
the beginning.  However, any computation with all qubits contributing
to the computation will generally become connected through a proper
assignment of simplicial complexes.

For a quantum computation, each of these $\bar{A}_{k}$ will, in
general, be a tensor product of single qubit Hilbert spaces. This
provides a simplified topology in which groups of points in the
original poset are consolidated into a single, local point in the
topology. The points then becomes a tensor product of Hilbert spaces
in the computation.  Instead, we wish to have points in the topology
to represent the individual, irreducible Hilbert spaces so as to
examine the connectivity of the computation at a given time for a
given thickness.

We generate the simplicial complex on individual Hilbert spaces with a
related approach that takes into account the unitary operators acting
in the computation.  We first supplement the poset of Hilbert spaces
with the poset of unitaries.  There exists a natural partial order
between the elements of the two distinct posets: (1) for every Hilbert
space ${\cal H}_{i}$ which is the domain of a unitary $U$, we have the
relation ${\cal H}_{i} \preceq U$.  Take a maximal anti-chain $A_{j}$
and (2) for every ${\cal H}_j$, i.e. each wire, which is in the image
of $U$ we have $U \preceq {\cal H}_j$.  Using the other partial order
relations on the two sets gives a partial order on the combined set.
The elements of $A_{j}$ form the vertices $v_{k}$ of the collection of
simplicial complexes to be generated.  Examine the next
foliation $A_{j+1}$ and the unitary operators acting on local subsets
of $A_{j}$ to generate local subsets of $A_{j+1}$.  We therefore can
introduce a partial order between unitary operators and Hilbert
spaces.  Recall the construction of the thickened anti-chain, which is
the set of points (unitaries for our purposes) in the poset meeting
the `thickening' criteria (\ref{eq:thickAC}).  The Hilbert spaces
immediately following the maximal unitaries in the thickened anti-chain
can be thought of as output Hilbert spaces of a composite unitary
whose domain is the anti-chain $A_{j}$.  The shadow of a given unitary
are thus the Hilbert spaces acting as the domain of the unitary.

A simplicial complex is constructed from the set $\bar{A}_{j}(U_{i})$
as the shadow of the unitaries $U_{i}$ on the anti-chain $A_{j}$;
\begin{equation}
\bar{A}_{j}(U_{i}) = \{  x \in  \J^{-}(U_{i}) \cap A_{j}\}.
\end{equation}
To each $\bar{A}_{j}(U_{i})$ we assign a complete graph or
$k$-simplex, with $k + 1 = \text{card}\left[\bar{A}_{j}(U_{i})\right]$.
For any two $\bar{A}_{j}(U_{i}), \bar{A}_{j}(U_{m})$ with non-empty
intersection, there is a simplex connecting them such that the two
complete graphs share the boundary given by the complete graph formed
through Hilbert spaces in their intersection.  This is guaranteed to
be a well-defined simplicial complex,\cite{Alexandrov:CombTopo}
analogous to the above the prescription.  This gives us a construction
for a simplicial complex in which the connectivity of the single qubit
Hilbert spaces is made explicit and is a consequence of the causal
properties of the computation.

%%%%%%%%%%%%%%%%%%%%%%%%%%%%%%%%%%%
%%%%%%%%%%%%%%%%%%%%%%%%%%%%%%%%%%%

\section{Quantum Cellular Automata}
\label{sec:qca}

Our interest now shifts to a particular model of computation: quantum
cellular automata (QCA).\cite{Zeilinger:QCA, Watrous:QCA,
 Schumacher:QCA, PDC:QCA} QCA are the natural extension of
classical cellular automata to the quantum regime.  However, a clear
understanding of QCA has not yet developed in sufficient detail to say
that there exists a complete theory of QCA.  Since our goal in
this manuscript is not to extract a particular meaning or property of
QCA, we will only need to specify the salient definitions and features
common to most known models of QCA in order to model a geometry of
the computation.  We can specify a particular model of QCA useful for
simulation and modeling the geometry.
\begin{mydef}
A QCA is a quadruple $({\cal R},\Sigma, {\cal\ N,\ U})$ consisting of a 
register ${\cal R} = \mathbb{Z}^{d}$, a finite collection of states $\Sigma$ for each
cell in ${\cal R}$, a collection of local neighborhoods, ${\cal N}$, and
a transition operator on the local neighborhoods, ${\cal U}$.\\
\end{mydef}

The register ${\cal R}$ labels the sites/qudits for our QCA and the
neighborhood specifies the dependencies of the QCA on some finite
collection of sites near an $x\in {\cal R}$.  The update rule acts on
the neighborhood of $x$, ${\cal N} \cup \{x\}$ to update the qudit at $x$.
For a given state of the neighborhood of a site, one operates on the
site with a single-qudit operator.  The total unitary on the
neighborhood is then given by $U( u_1, u_2, \ldots, u_{N})$ where $N$
is the number of states accessible by the neighborhood and the $u_i$
are single qubit operations. We thus consider the update rules as
multiply-controlled unitaries with target $x$.  This feature will be
key for considering the topology of the QCA at a given time.

\subsection{Block-Partitioned QCA}
\label{subsec:BPQCA}
Given a infinite register ${\cal R}$, it is unclear as to how one
defines appropriate local updates while ensuring unitarity of the
global system.\cite{Schumacher:QCA, PDC:QCA}  A clear way avoid this is to implement
partitioned QCA.
\begin{mydef}
  A partitioned QCA is a 6-tuple $\{ {\cal R}, S, {\bf B}, \Sigma,
  T , {\cal U}\}$ where ${\cal R}= \mathbb{Z}^{d}$ is the register,
  $S$ is a sublattice of ${\cal R}$, ${\bf B}$ is a partitioning
  scheme on ${\cal R}$, $\Sigma$ is a finite set of cell states, $T$
  is the local period for a global update, and ${\cal U}$ is the local
  update rule. \\ \\
\end{mydef}
An example of a partitioned QCA is the block partitioned
QCA (BPQCA). \cite{Brennen:QCA} On the register of sites, the qubits, the
partitioning of the register is done according to species of qudits
and the assignment of a neighborhood to each species.  If one takes two
species $A$ and $B$ and a nearest neighbor neighborhood, then a global
update rule can be defined as the successive application of an update
rule on the species $B$ qudits (with their neighborhood) followed by
the update on species $A$ qudits.

To further simplify the computation (and more accurately model
physical implementations), we will supplement the block-partitioning
with boundary conditions.  In a finite $d$-dimensional BPQCA, the
register is given by ${\cal R} = \mathbb{Z}^{d}_N$ with static
boundaries given by ancillae qudits.  The neighborhoods of sites near
the boundary can then be reduced to just those sites within ${\cal
  R}$.  In the rest of this manuscript, we will exclusively examine the geometry
of such a block-partitioned QCA with sites containing qubits.

\subsection{1D QCA}
\label{subsec:1dgeom}
\begin{figure}[b]
  \begin{center}
    \begin{tabular}{c}
      \includegraphics[width=3.5in]{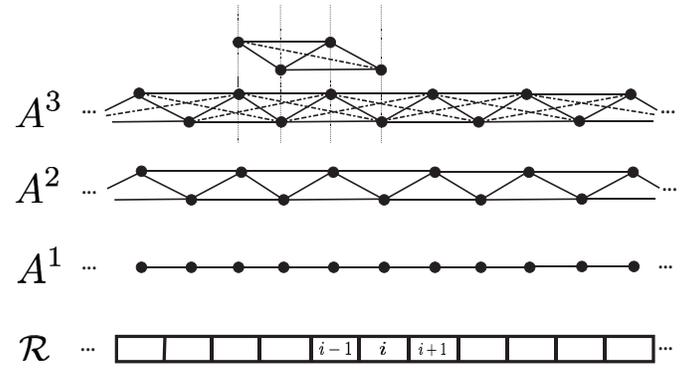}
    \end{tabular}
  \end{center}
  \caption[example]
  { \label{fig:1dComplexes} The simplicial complexes generated on a 1D
    BPQCA generated by thickened anti-chains.  The thickness of the
    anti-chain corresponds to the number of local update rules
    applied, where $A$ and $B$ species updates are distinct.  We see
    that as we increase the thickened anti-chain generating the
    complex, the global topology remains essentially constant,
    however, locally the system generates higher-dimensional
    cross-sections.  Since there is no significant change in the
    topology as thickness increases, we can take the earliest
    simplicial complex in the stable regime of the filtered homology,
    $A_1$. This will represent the structure identifying neighboring
    qubits and determine how we specify information distances. }
\end{figure} 

Suppose now that ${\cal R} = \mathbb{Z}_{N}$ and each site is a
two-level system, i.e a qubit.  We have a 1-dimensional BPQCA of
$N$-qubits with two boundary qubits on the two ends of ${\cal R}$.  We
define the neighborhood of a site $i$ as the set of nearest neighbors,
${\cal N}_{i} = \{i-1, i+1\}$.  We block partition ${\cal R}$ into two
species, even (A) and odd (B) following Brennen, {\em
  et. al}.\cite{Brennen:QCA} The update rules are given by four
unitary operators $u_{0}, u_{1}, u_{2}, u_{3} \in {\rm SU}(2)$.  The
subscript on these single-qubit operations labels the basis vectors
for the Hilbert space of the qubits neighboring the site being
updated, i.e. $0\rightarrow |00\rangle,\ 1\rightarrow |01\rangle,\ 
2\rightarrow |10\rangle,\  3\rightarrow |11\rangle$.  The global update
 is the combined update $U = U^{A}(u_0, u_1, u_2, u_3) U^B(u_0,
u_1, u_2, u_3)$.  The update rule simplifies to $U^0(u_{0}, u_1)$ and
$U^{(N-1)}(u_0, u_2)$ on the sites at the boundary of ${\cal R}$.

What is the topology generated by the evolution of this system?  How
does one measure the geometry?  Let us examine first the update rule.
Since the global update is decomposed into two distinct operations on
${\cal R}$, we examine each update as an individual time-step of the
system.  Following the arguments of Sections~\ref{subsec:causal} and
\ref{subsec:fol2geom}, we generate simplicial complexes based on the
causal future of the slice in the computation.  In the causal past of
single local update, we have the local neighborhood ${\cal N}\cup
\{i\}$.  In general, this will be mapped to a 2-simplex.  The
simplicial complex of the first non-trivial thickened anti-chain then
becomes a chain of 2-simplexes with the edge $\overline{i, i+1}$
shared by two such 2-simplexes.  Any edge $\overline{i, i+2}$
is contained in only one 2-simplex.  As we increase the thickness and
derive the simplicial complex on the initial slice, the effect is to
add two-dimensions to the individual building blocks while retaining
the global 1-dimensional nature, i.e. the system acts as a cylinder
with a $2t$-dimensional cross-section.  The infinite QCA is
effectively 1-dimensional with local structure generated by
interaction via the light cones.

If we utilize the fact that each local update rule $U^i$ acts on
${\cal N}\cup \{i\}$ as a multiply-controlled quantum gate, then we
simplify the model and preserve the connected nature of the 1d BPQCA.
We note that the controlled nature of the update rule is such that
after a single update is applied, two sites, $i$ and $i+2$ have no
causal influence on one another.  They act simply as controls for site
$i+1$.  Hence, in modeling the causal behavior of the site, the
simplicial complex incorporating the controlled-nature of the update
removes the edges $\overline{i, i+2}$.  The simplicial complex induced
by a single update yields the topology of a discrete line.  Increasing
the thickness of the thickened anti-chain induces a topology on the
slice that increases the dimensionality of the cross-sectional
surfaces along the longitudinal axis, creating a cylinder of
cross-sectional dimension $d>0$ (Figure~\ref{fig:1dComplexes}). The
topology is thus stable with a topology that is analogous to that of
${\mathbb R}$.  It is consistent to take as the simplicial complex,
the initial complex which is simply connected.

This construction yields a topology endowed with a local
metric that has at most one local degree of freedom.  The intrinsic
geometry has no curvature and is entirely characterized by the local
scale factor.  We therefore specify the entire intrinsic geometry by
giving the local distance between any two points in the topology.  In
the 1D BPQCA, this intrinsic geometry measures only bipartite
correlations between nearest neighbors and carries no information
about more distributed measures of farther separated correlations.   

\begin{figure}[t]
  \begin{center}
    \begin{tabular}{c c c}
      \includegraphics[height=1.25in]{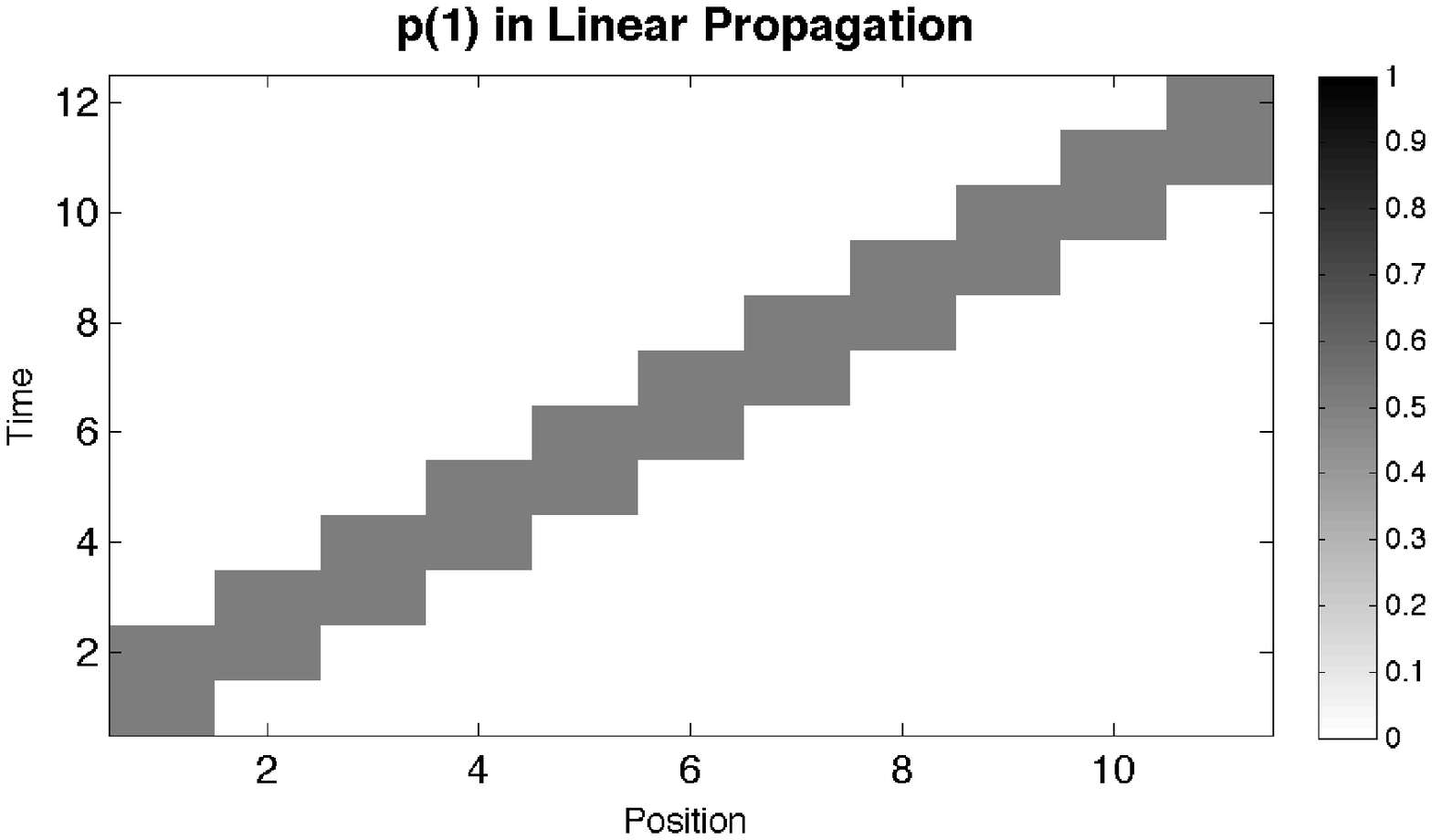}  &&  \includegraphics[height=1.25in]{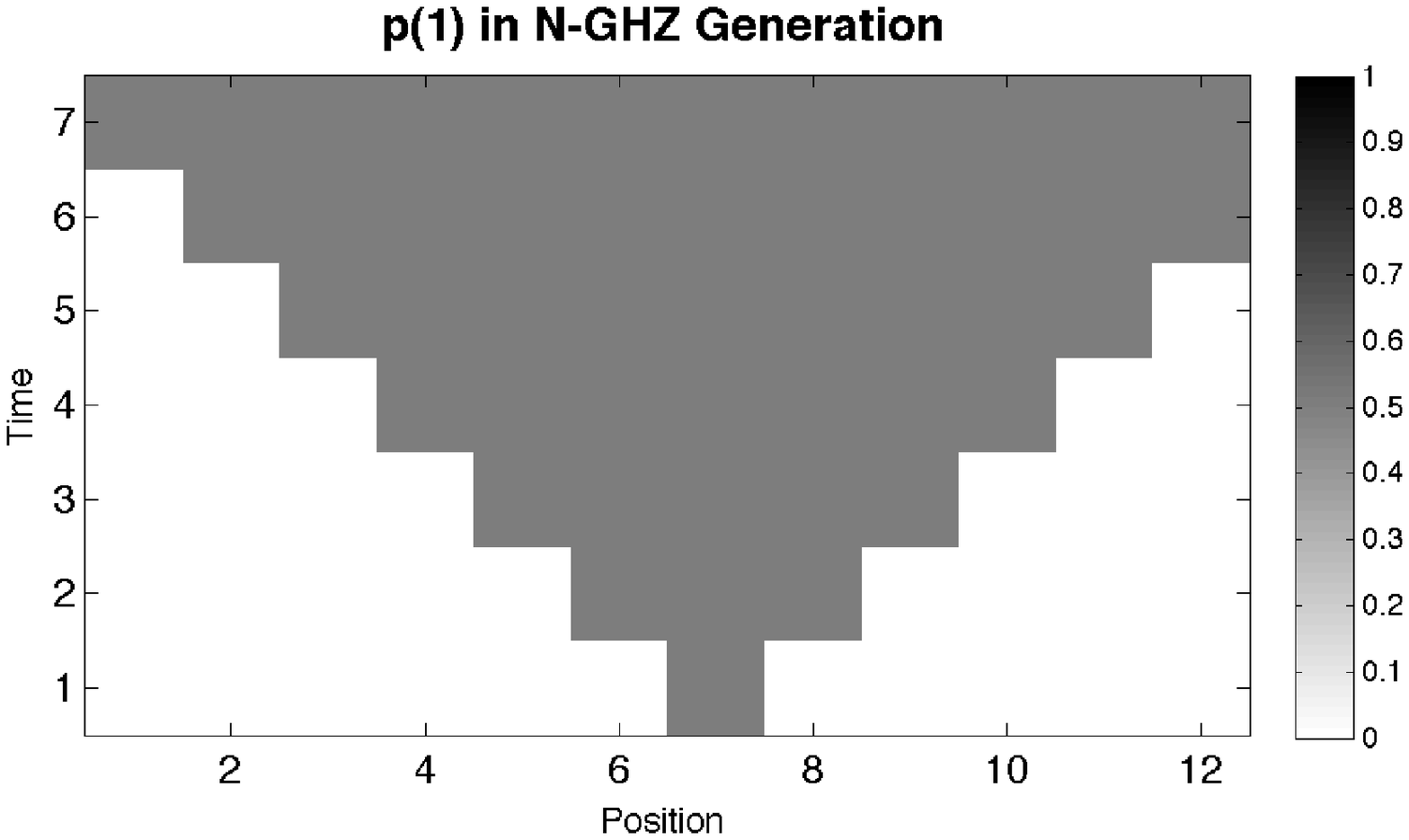} \\
      (a) &  & (b)\\
      \includegraphics[height=1.25in]{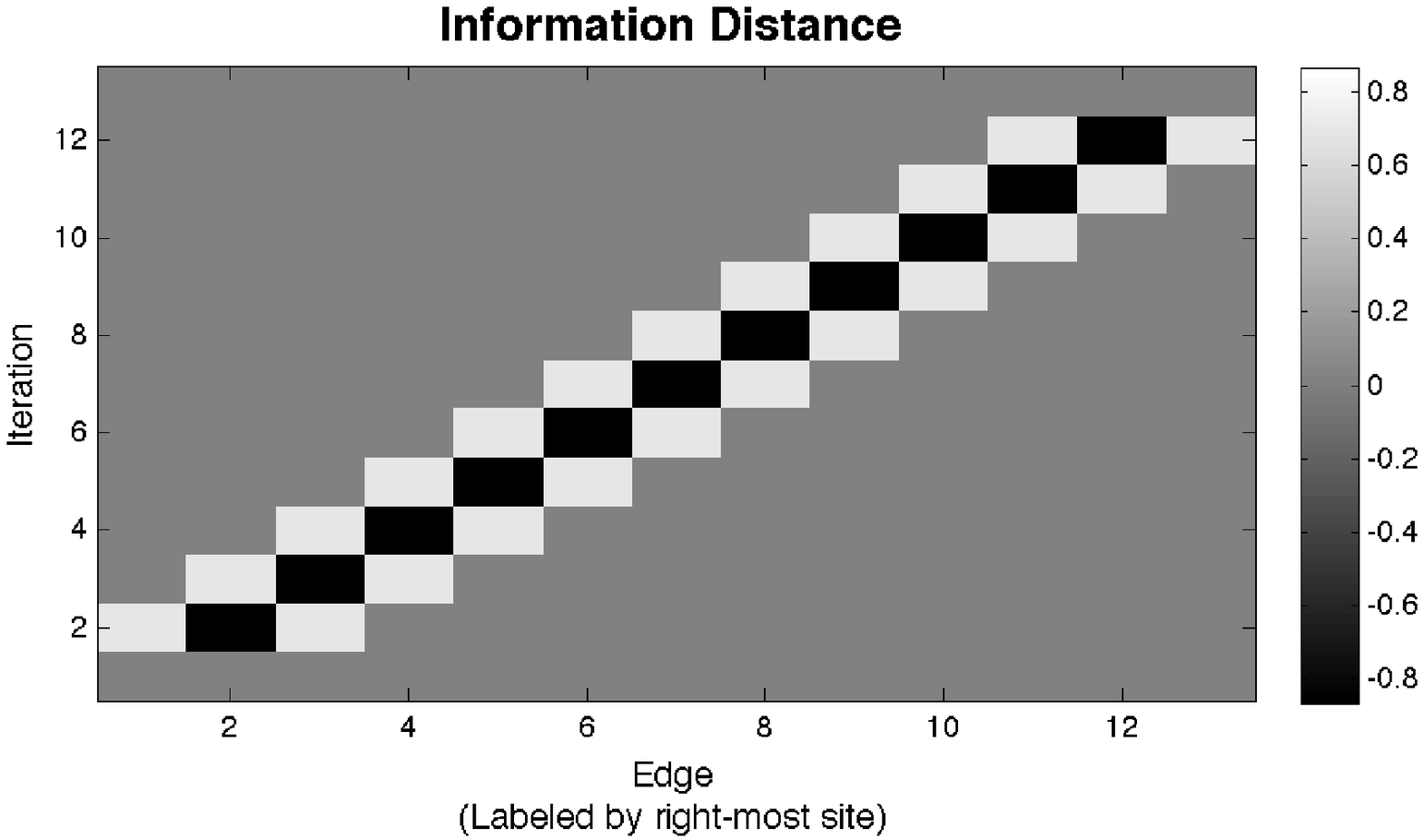}  &&\includegraphics[height=1.25in]{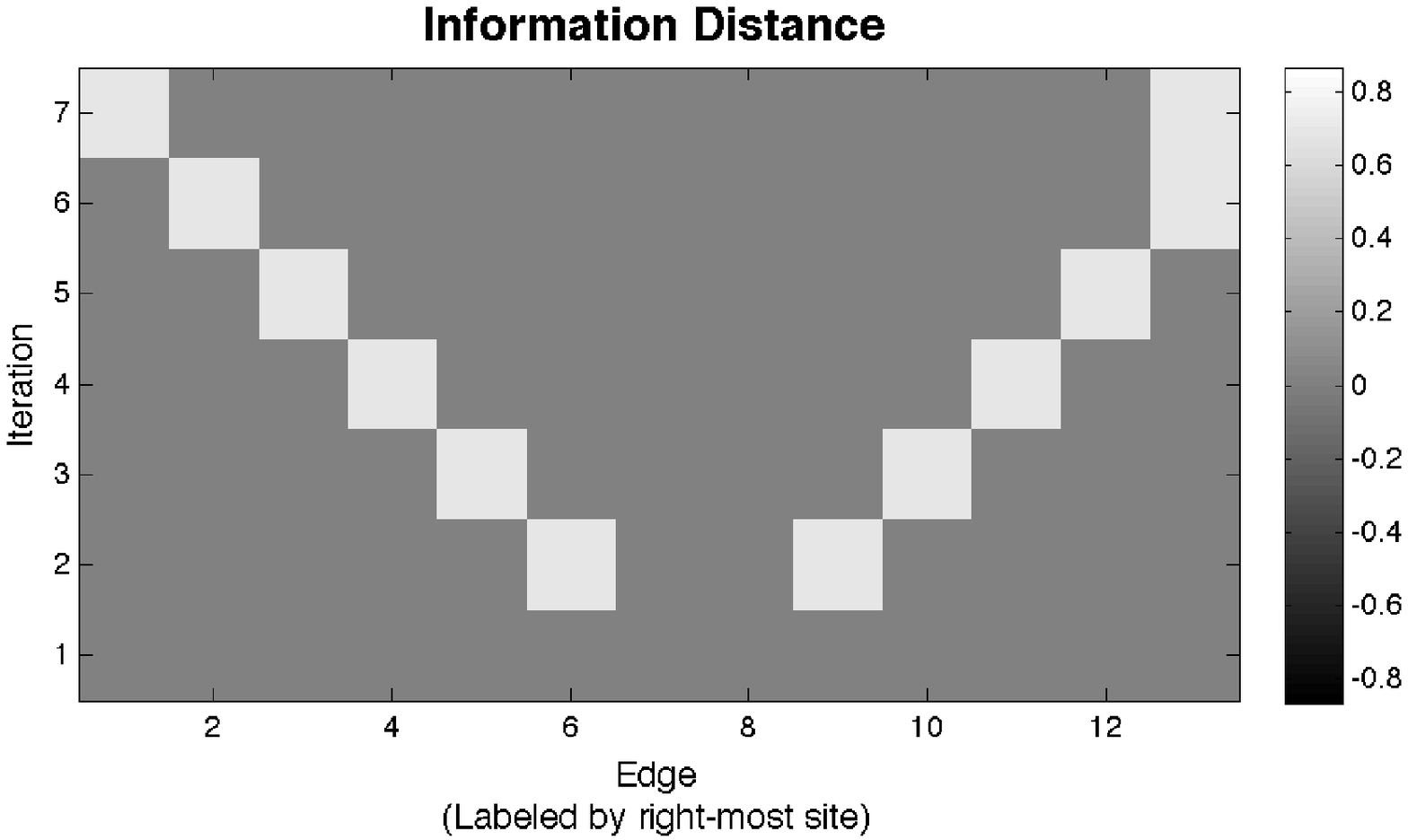}\\
      (c) & & (d)
    \end{tabular}
  \end{center}
  \caption[example] 
  { \label{fig:1dLP-GHZ} (a) The probability, $p(1)$, that the qubit
    is in the state $|1\rangle$ for the propagation of a single qubit
    across the 12 qubit QCA.  The color scale is such that black
    corresponds to $1$ and white corresponds to $0$.  (b) The
    probability, $p(1)$, for generation of the 12-qubit GHZ state. (c)
    Information distances in the linear propagation of a qubit across
    the 1D BPQCA.  We explicitly show the distances between qubits in
    the QCA and the fixed boundary qubits in this plot. Negative
    distances (black) correspond to bipartite entanglement, while
    white/light grey corresponds to positive information distances.
    (d) Information distances in GHZ generation. We see no evidence no
    bipartite quantum (negative distance) correlations.  However,
    positive information distances form a boundary around the GHZ
    state.  This positive distance boundary is key to identifying the
    form of entanglement present in the system.  }
\end{figure}

In some standard simulations of 1D BPQCA,\cite{Brennen:QCA} we
demonstrate (Figure~\ref{fig:1dLP-GHZ}) propagation of an unknown
qubit across the chain or generation of a GHZ state.  We then generate
distance information from the Zurek distance on the intrinsic geometry
of the QCA. The register for the computation is the set of qubits
${\cal Q}= \{q_1, \ldots, q_{N}\}$ with boundary qubits ${\cal B}=
\{q_0, q_{N+1}\}$.  These qubits are partitioned into even (B) and odd
(A) species.  The single species update rule is given by $U^{A(B)}:=
U^{A(B)}(\mathds{1}, e^{-i\frac{\pi}{2}\sigma_{x}},
e^{-i\frac{\pi}{2}\sigma_{x}}, e^{-i{\pi}\sigma_{x}} )$.  Propagation
of an unknown state $|q_{1}\rangle = |\psi \rangle$ is achieved by
$N/2$ successive applications of $U = U^{A}U^{B}$ and a final
$Z$-rotation on $1_{N}$.  In Figure~\ref{fig:1dLP-GHZ}(a, c) we show
results from the simulation on a 12 qubit register.  It is interesting
to note that if the update rule is applied to the seed $|{\cal
  Q}\rangle = |100\cdots 0\rangle$, then no entanglement is generated.
With an update rule and an initial seed in the same basis, the state
stays separable and the information distance is 0 throughout the
computation, i.e. the geometry of the computation is a point.

To generate an $N$-GHZ state, 
$$|N\text{-GHZ}\rangle = \frac{1}{\sqrt{2}}
( |\underbrace{00\cdots 0}_{N-\text{times}}\rangle
+|\underbrace{11\cdots 1}_{N-\text{times}}\rangle ),$$ we follow the
procedure above and seed $q_{N/2}$ ($q_{N/2+1}$) with $|+\rangle$ when
$N \text{mod} 4 = 0$ ($N \text{mod} 4 = 2$).  The local, single-step
update rule is again given by $U^{A(B)}:= U^{A(B)}(\mathds{1},
e^{-i\frac{\pi}{2}\sigma_{x}}, e^{-i\frac{\pi}{2}\sigma_{x}},
e^{-i{\pi}\sigma_{x}} )$ and is applied over $k = N/4$ ($ k= (N-2)/4$)
iterations on both the $A$ and $B$ species, i.e. $k$ global iterations
or $2k$ single-step iterations.  When $N\text{mod}4 = 2$, one must
apply an additional single-step update on the $B$ species.  The final
$N$-GHZ state is obtained after a single-qubit phase correction,
$e^{(-1)^{k}i\frac{\pi}{4} \sigma_{z}}$.  If we examine the geometry
generated from this evolution, we find that any given slice yields no
specific revelation as to the entanglement in the system.  However,
coupling together the intrinsic geometry of the slice with either the
extrinsic geometric information (non-nearest neighbor information
distances) or knowledge of the evolution of the system tells gives
indications as to the structure of entanglement found in the system.

From the non-nearest neighbor distances we obtain a distance matrix of the form
\begin{equation}
\left(
\begin{array}{c c c c c c c c c}
0 & 0 & 0 & d_{1i_1}&  \cdots & d_{1i_n} & 0& 0 &0 \\
0 & 0 & 0 & \vdots&  \ddots & \vdots & 0& 0 &0 \\
0 & 0 & 0 & d_{(i_1-1) i_1}&  \cdots & d_{(i_1-1) i_n} & 0& 0 &0 \\
d_{i_1 1} & \cdots & d_{i_1 (i_1 -1)} & 0 & \cdots & 0 &d_{i_1 (i_n+1)} & \cdots & d_{i_1 N}  \\
\vdots & \ddots & \vdots &  \vdots & \ddots &  \vdots  &\vdots& \ddots &\vdots \\
d_{i_n 1} & \cdots & d_{i_n (i_1 -1)} & 0 & \cdots & 0 &d_{i_n (i_n+1)} & \cdots & d_{i_n N}  \\
0 & 0 & 0 & d_{(i_n+1)i_1}&  \cdots & d_{(i_n+1) i_n} & 0& 0 &0 \\
0 & 0 & 0 & \vdots&  \ddots & \vdots & 0& 0 &0 \\
0 & 0 & 0 & d_{N i_1}&  \cdots & d_{N i_n} & 0& 0 &0 \\
\end{array}
\right)
\end{equation}
where $d_{ij}>0$ and the string of qubits $\{ q_{i_1} , \ldots
q_{i_n}\}$ is dependent on the global iteration number of the QCA.
Since the global state is pure, the existence of non-zero distances
indicates non-local correlations in the system.  The structure of the
distance matrix demonstrates mutlipartite entanglement on the qubits
in a block whose information distances are all null.  This is clear
since the distance between any qubit in the interior block and any
qubit outside the block is positive, the distances inside the block are
all null, and the distances between qubits entirely outside the block
are all 0.  We need only identify whether the interior block or
exterior block contains multipartite, non-local correlations.  Since
on the first iteration, $q_1$ cannot have causal influence or be
causally influenced by qubits outside its neighborhood, we must find
that the interior block exhibits non-local correlations.  Extending
this from iteration 1 to iteration $k$ allows us to determine the
location of multi-partite non-local correlations at any instance in
the QCA for this example. 

If we instead wish to keep the focus on the intrinsic geometry and the
causal evolution, we again can make certain observations about the
nature of non-local correlations in the system.  We recognize that
causal influence has a finite velocity in the system and that the
system starts in a separable state.  The state at any iteration $j\geq
1$ has the general property that there are three regions within which
the information distances are all null.  We can simplify this to two
regions, the exterior and interior in an obvious way. We assume we
know only the information distances and not necessarily the state of
the system at any iteration. As we iterate, we have two options: (1)
The two regions separated by the positive distance boundary are
regions of separable states (with possible non-local correlations
between them), or (2) one of the regions is a set of qubits exhibiting
multipartite non-local correlations.  We can rule out the first
immediately as the global system is pure and at the first iteration
non-local correlations between the two subsets of qubits would be
impossible given the update rule.  On the latter option, any region
containing $q_{1}$ would then require non-local correlations between
$q_1$ and a qubit outside its local neighborhood.  As before, this is
a scenario that is impossible given the finite velocity of causal
influence in the QCA.  Applying this to either of the exterior
regions and following the evolution tells us that there must be
multipartite, non-local correlations on the set $\{q_{i_1}, \ldots,
q_{i_n}\}$.

\begin{figure}[t]
  \begin{center}
    \begin{tabular}{c  c c c }
      \includegraphics[width=1.5in]{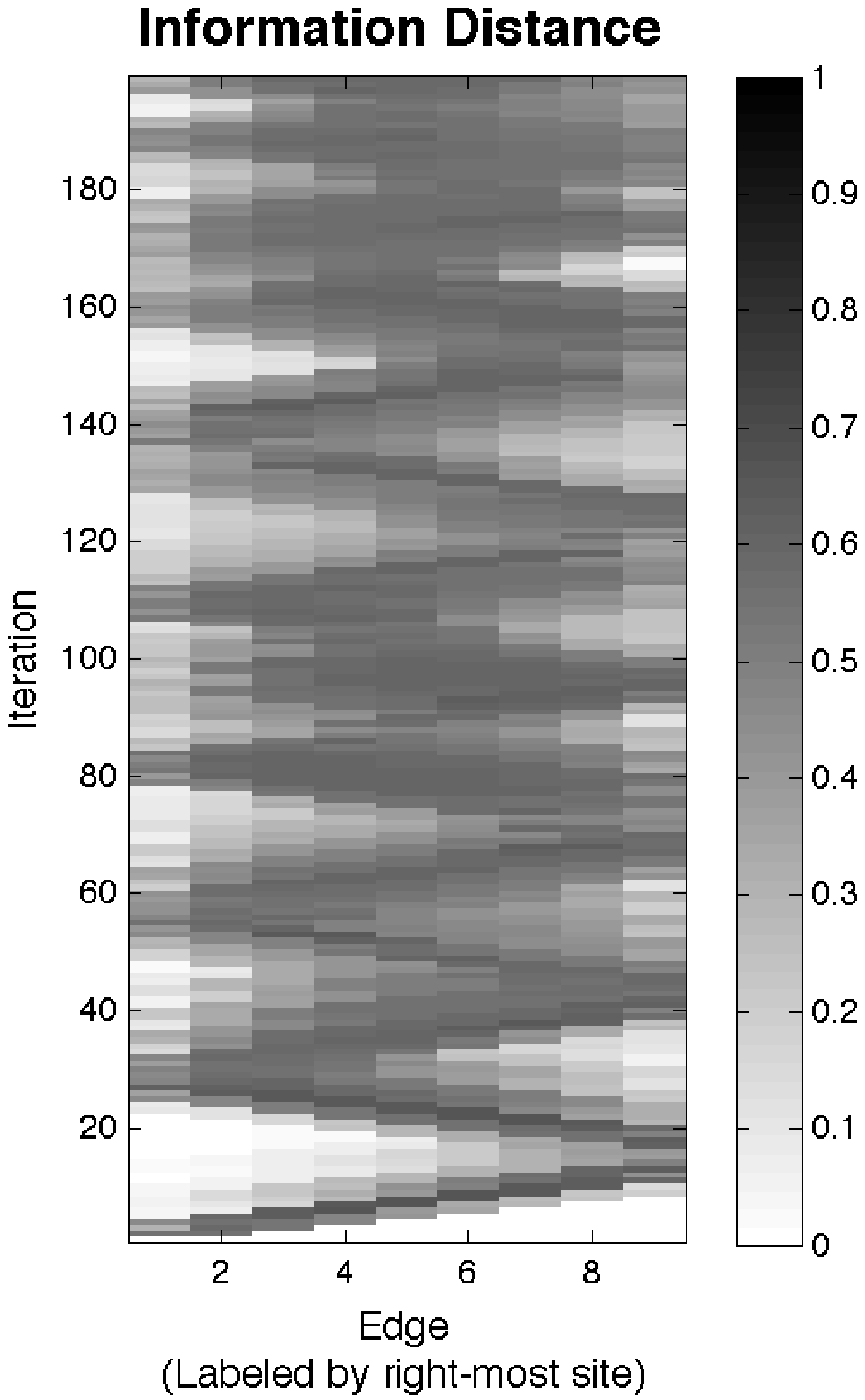}  & \includegraphics[width=1.5in]{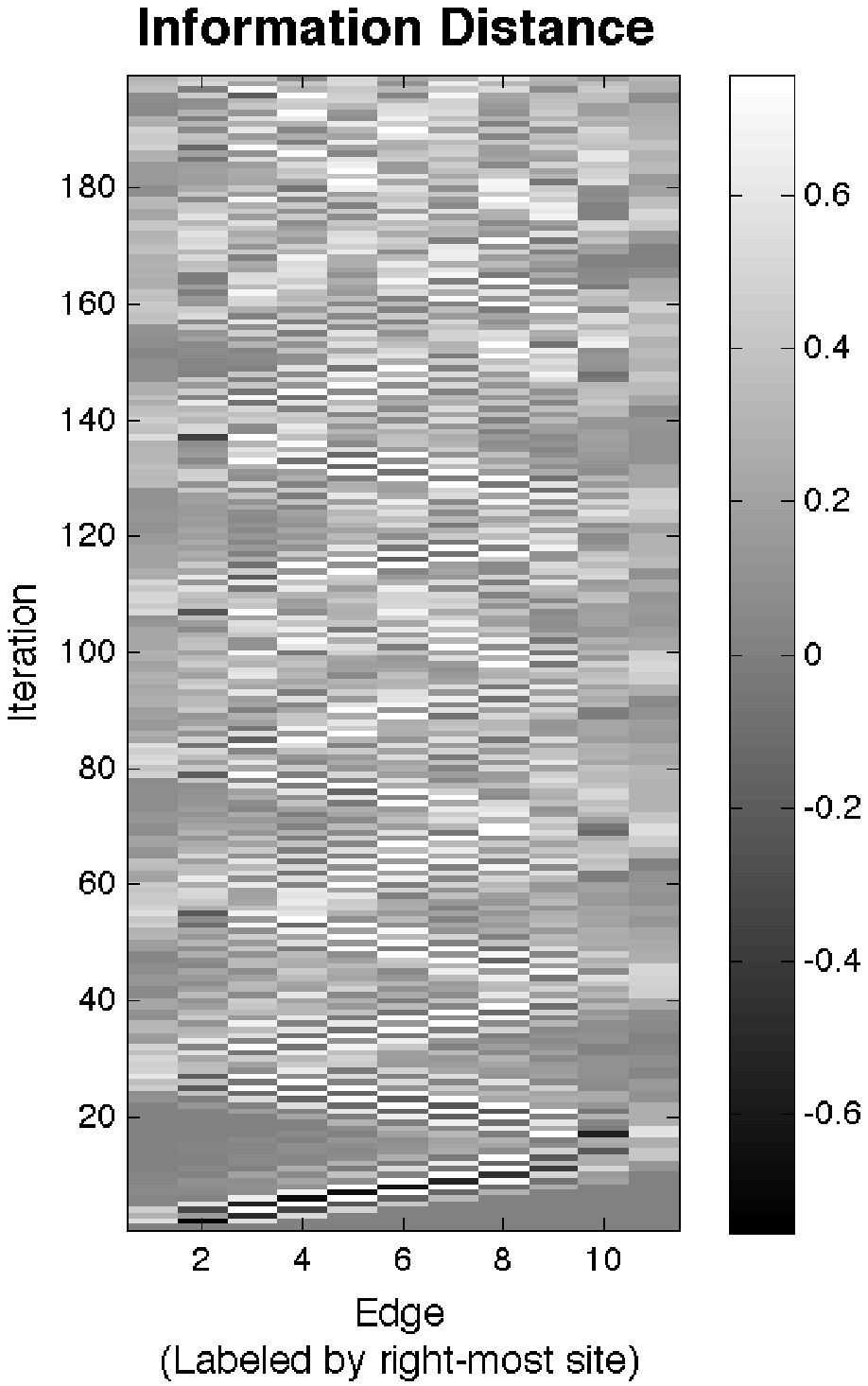}   
      & \includegraphics[width=1.5in]{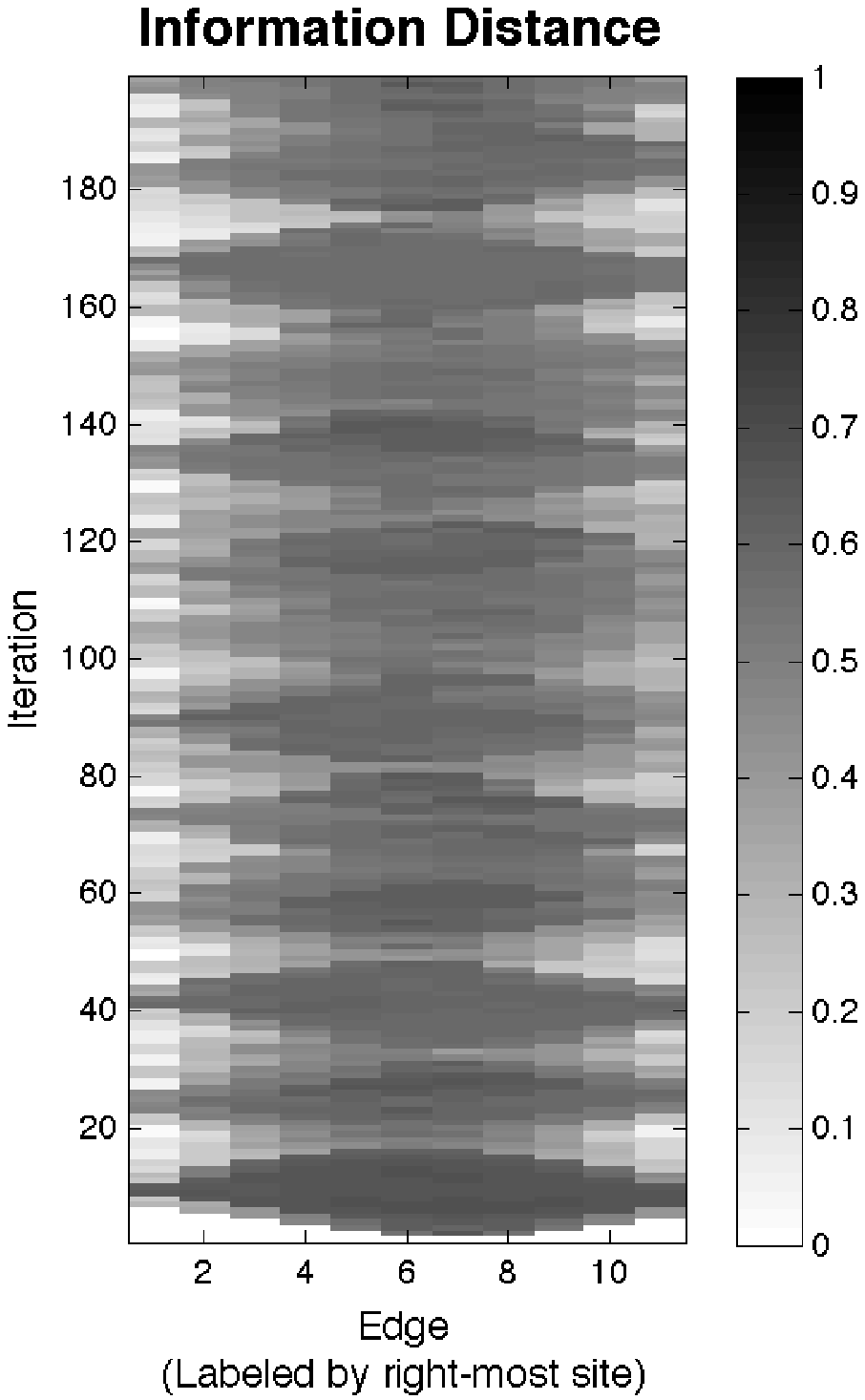}  & \includegraphics[width =1.5in]{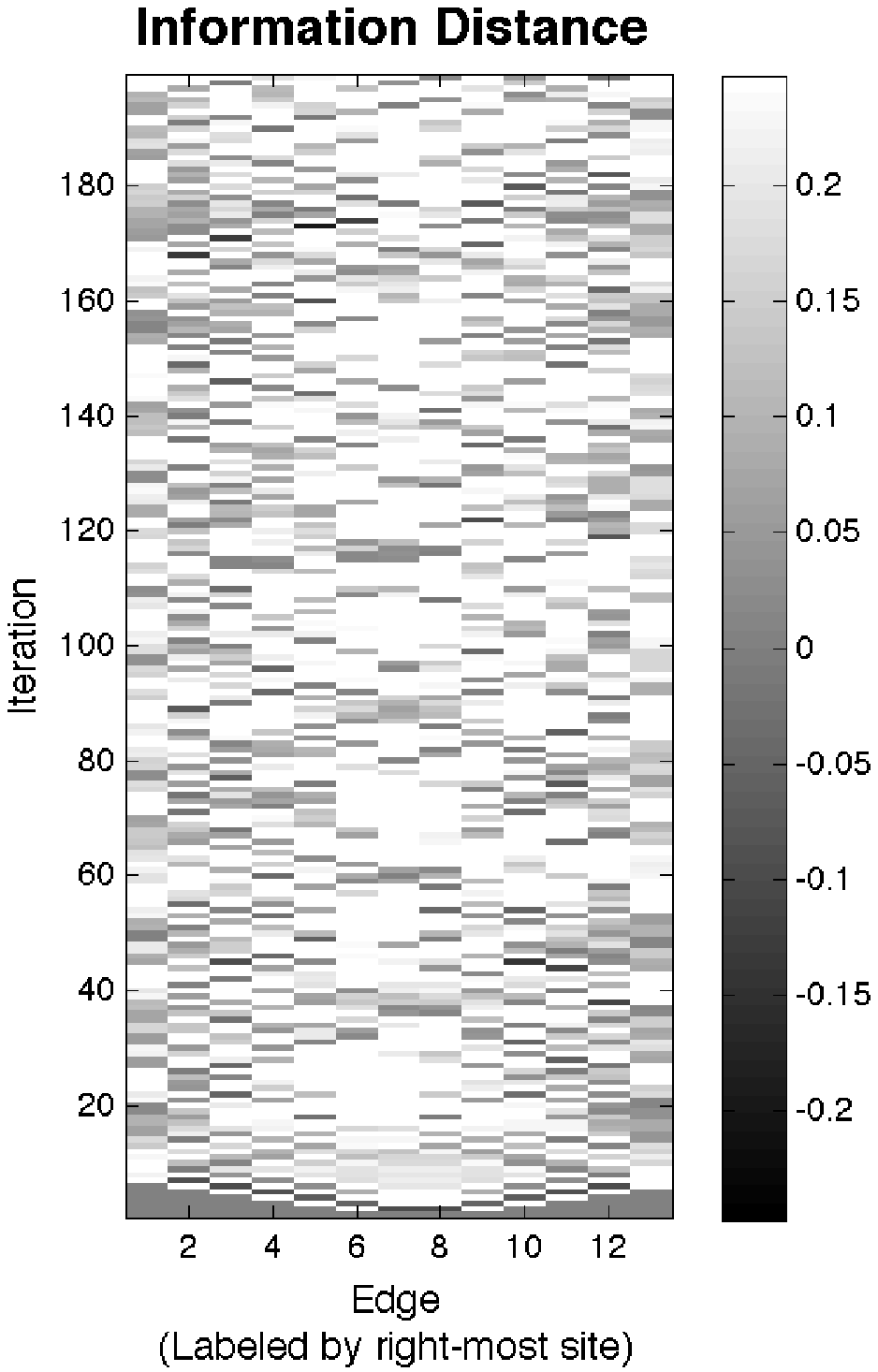}\\
      (a) &(b) & (c) & (d)
    \end{tabular}
  \end{center}
  \caption{ \label{fig:Pi3} The QCA with update rule
    $U^{A(B)}(\mathds{1}, e^{-i\frac{\pi}{3}\sigma_{x}},
    e^{-i\frac{\pi}{3}\sigma_{x}}, e^{-i\pi\sigma_{x}} )$. (a) The
    reduced entropy, $S(q_i)$, obtained through partial trace over
    ${\cal Q}-q_{i}$.  The QCA of 10 qubits is seeded with qubit
    $|q_{1}\rangle = |+\rangle$.  We see that $S(q_i)$ is always
    positive on the sites.  (b) The information geometry shows
    distinct regions of bipartite entanglement.  Bipartite
    entanglement is captured by negative information distances (darker
    regions) in the geometry.  Lighter regions correspond to the
    positive information distances in which classical correlations are
    most prevalent.  While it is possible to identify entanglement via
    negative information distances, positive distances in the geometry
    do not necessarily correspond to lack of entanglement.  (c) The
    reduced entropy on the QCA with 12 qubits seeded with qubit
    $|q_{\frac{N}{2} +1}\rangle = |+\rangle$.  (d) The information
    distances on the same system as (c).  Again local, bipartite
    quantum correlations propagate and diffuse as the QCA
    progresses. However, we can identify instances in the computation
    where nearest neighbors exhibit such quantum
    correlations. Extending information distances to geometric content
    beyond the intrinsic topology can lead to richer characterization
    of the geometry and entanglement.}
\end{figure}

We now consider an example where the bipartite entanglement is not so
straightforward. We take as the single-step update rule
$U^{A(B)}(\mathds{1}, e^{-i\frac{\pi}{3}\sigma_{x}},
e^{-i\frac{\pi}{3}\sigma_{x}}, e^{-i\pi\sigma_{x}} )$. The register is
seeded with $|q_{i}\rangle = |+\rangle$ and all other $|q_{j}\rangle
=|0\rangle$ ($j\neq i$).  In this case, the update rule generates some
entanglement between pairs, and distributes the entanglement as the
computation goes forth.  Figure~\ref{fig:Pi3} shows both the reduced
entropy and the information distance for this update rule.  Non-zero
reduced entropy of a qubit $q_{k}$ in a system where the global state
is pure (and hence has zero entropy) can indicate the presence of
entanglement between the reduced system and the state of ${\cal
  Q}-q_{k}$.  However,the nature of the entanglement is not clear.
Instead, the information distance in the intrinsic geometry gives us
the ability to identify the existence of local (nearest neighbor),
bipartite entanglement.  The negative information distances correspond
to the necessary presence of quantum, non-local correlations, and the
positive distances still provide a measure of the existence of
entanglement between a qubit and some distant subset of the QCA not
immediately connected to it.

From the local nature of the simplicial complex, we see that the
information distance as a measure of the intrinsic geometry can only
tell us about the correlations between nearest neighbors.  In cases
where the information distance is negative, we have direct indication
of quantum, non-local correlations.  We have already discussed how to
utilize either the extrinsic geometry or causal content to identify
multipartite non-local correlations that still may lead to
non-negative distances.  However, this was specific to generation of
the GHZ state.  In more general cases, we could relax the assumption
of taking the earliest simplicial complex in the stable regime of the
filtered (with respect to the thickened anti-chains) homology.  We
could instead use the simplicial complexes with thickened anti-chains
built into the past of the anti-chain $A_j$.  It would be natural then
to take the thickness equal to the number of iterations performed.
This would amount to an accumulation of the past light cone into a
mutually interacting simplex in the cross section of the topological
cylinder, $S^{d} \times {\mathbb{R}}$.  An alternative is to couple
the intrinsic geometry with an embedding in the space constructed from
information distances between all bipartite decompositions of all
possible subsets in the system.  This is not favorable as the
parameters in this type of model scales exponentially with the number
of qubits.

%%%%%%%%%%%%%%%%%%%%%%%%%%%%%%%%%%%
%%%%%%%%%%%%%%%%%%%%%%%%%%%%%%%%%%%
\section{Discussion}
\label{sec:disc}

We have proposed here a new view of quantum computation based on the
configuration of qubits at any given time of the computation. Central
to this proposed perspective is that one of the essential features of
the computation is not the exact states of the qubits in the
computation, but rather the information content in the configuration
of qubits. A pure, separable state of $N$-qubits contains no more
information than if the qubits are set in the computational state
$|00\cdots 0\rangle$, since local unitaries applied to the individual
qubits that rotate each qubit into the zero state do not change the
information content.  Indeed, there is no physical distinction between
an arbitrary, pure, separable state and the state $|00\cdots
0\rangle$, only a local basis change.  We see this in the information
distance as the total absence of any geometric content in such a
state. The ability of a computation to yield a
given output is contained in the correlations in the constituents of
the geometry. To give the particular state of the computation is akin
to coordinatizing the geometry. This situation can be likened to that
of general relativity in which one is perfectly allowed to ask the
position or state of a particle/system with respect to some
coordinates or even some other reference system. However, in general
relativity it is not reasonable to ask of the physical system the {\em
  absolute} position or momentum of a particle/system.  Likewise, in
quantum computation when we put a set qubits into the computational
basis, we have specified a coordinatization on the states.  This
coordinatization does not carry in it any physical content.

This is not a view that is incompatible from the view proposed by
Nielsen, {\em et. al.}\cite{Nielsen:QC2006, Nielsen:QC2008} where the
emphasis was on the unitary operators in the computation. There, one
need not specify an initial or final state as it is not necessarily
relevant to the computation. Rather, the relevant content is contained
in how the information is distributed across the qubits and how the
system has changed from the initial to the final states.  

In a quantum system one may only ask questions with respect to a
certain type of measurement.\cite{Wheeler:itfrombit} Without such
specification, it is impossible to make heads or tails of the qubits
in the computation.  We only extract information from the computation
by composing a string of bits based on the yes/no questions with
respect to a given basis for each qubit.  Without such
coordinatization, the state of the string of qubits has given no
direct answer.  The ability to extract the result of computation is
only realized when we know what questions to ask, i.e. when we know
with respect to which coordinization the system is to be evaluated.
Of course, even then, the system will not have a unique state from
which to extract the answer.  The result of the quantum computation is
the {\em in toto} information encoded in the qubits, independent of
local unitary transformations.  This is similar to the situation one
must face in general relativity where the physical content is encoded
in quantities invariant under local $GL(4)$ transformations that can
modify the local basis, the metric components, {\em etc.}

This present manuscript has outlined this approach and provided some
initial analysis on 1D QCA. The analysis of non-local correlations in
QCA shows some functional possibilities of generating the
computational geometry. Foremost is the ability to identify bi-partite
entanglement. We have shown explicitly the appearance of negative
information distances as a way to identify quantum correlations.
However, negative distances are only sufficient and not necessary for
there to be quantum correlations present in a general mixed state. A
complete classification of entanglement in a quantum computation (on
pure states) requires either a deeper analysis on the decomposition of
information distance into classical and quantum components and/or the
distance measures between all possible bipartite pairs in the quantum
system.  For classification of entanglement in computations on mixed
states, it is necessary to examine the decomposition of the
information distance into classical and quantum components along the
lines of Vedral, {\em et. al.}\cite{Vedral:1997}

The application of this approach to 1D QCA shows some utility in
identifying topologically local quantum correlations.  As we extend
this approach to higher-dimensional cases, the geometric content of
the distance information imposed on the topological scaffolding
becomes much more rich.  Beyond information distance, it will be
necessary to understand quantities such as information areas or
volumes. The intrinsic topology of the computation at any given time
was constructed so as to identify paths of information flow as the
computation moves forward.  Higher dimensional analogs of this notion
will aim to indicate the correlations between many qubits who share
mutual influence.  Higher dimensional examples will additionally allow
us to incorporate curvature as way to study the information flow
through the system.  Recent work in information flow on networks has
suggested that the curvature gives an indication of the load-balancing
or congested nature of the information flow.\cite{Jonckheere:NetCurv}

With the notion that curvature gives indication as to load-balancing
or congestion in an information carrying/processing system, one way to
optimize a quantum computation is to distribute the computation across
the quantum register.  Optimally distributing the information
processing will ensure maximum use of quantum parallelism,
i.e. maximal utilization of the degrees of freedom available to the
register.  Such optimization can, in principle, be achieved by
applying geometric gradient flows, e.g. Ricci flow\cite{Hamilton:RF},
act as diffusive flows for curvature.  Future work will focus on
extending this current approach to higher-dimensional models that
allow for such an analysis on the curvature in the computation.

%%%%%%%%%%%%%%%%%%%%%%%%%%%%%%%%%%%%%%%%%%%%%%%%%%%%%%%%%%%%%
\acknowledgments    

JRM was supported through a National Research Council Research
Associateship Award at AFRL Information Directorate for this research.
PMA and JRM acknowledge the partial support of the Air Force Office of
Scientific Research (AFOSR) for this work. HAB acknowledges support
from AFRL Information Directorate under grant FA 8750-11-2-0275.  JRM,
PMA, and HAB would like to thank Warner A. Miller for numerous helpful
discussions. Any opinions, findings, and conclusions or
recommendations expressed in this material are those of the authors
and do not necessarily reflect the views of AFRL.

\bibliography{QCAgeom}   %>>>> bibliography data in report.bib

\begin{thebibliography}{10}

\bibitem{Nielsen:QC2006}
Nielsen, M.~A., Dowling, M.~R., Gu, M., and Doherty, A.~C., ``Quantum
  computation as geometry,'' {\em Science}~{\bf 311}(5764),  1133--1135 (2006).

\bibitem{Nielsen:QC2008}
Gu, M., Doherty, A., and Nielsen, M.~A., ``Quantum control via geometry: An
  explicit example,'' {\em Phys. Rev. A}~{\bf 78},  032327 (2008).

\bibitem{Blute:2003}
Blute, R.~F., Ivanov, I.~T., and Panangaden, P., ``{Discrete Quantum Causal
  Dynamics},'' {\em Int. J. Theor. Phys.}~{\bf 42}(9),  2025--2041 (2003).

\bibitem{Zurek:1989dist}
Zurek, W.~H., ``{Thermodynamic cost of computation, algorithmic complexity and
  the information metric},'' {\em Nature}~{\bf 341}(6238),  119--124 (1989).

\bibitem{Schumacher:1991dist}
Schumacher, B., ``Information and quantum nonseparability,'' {\em Phys. Rev.
  A}~{\bf 44}(11),  7047--7052 (1991).

\bibitem{Feynman:QC1982}
Feynman, R.~P., ``{Simulating physics with computers},'' {\em Int. J. Theor.
  Phys.}~{\bf 21}(6/7),  467--488 (1982).
\newblock reprinted in [{\em Feynman and Computation}], A.J.G. Hey (ed.),
  Perseus Books: Reading, MA (1999).

\bibitem{Zurek:Discord}
Olliver, H. and Zurek, W., ``{Quantum Discord: A Measure of the Quantumness of
  Correlations},'' {\em Phys. Rev. Lett.}~{\bf 88}(1),  017901 ( 4 pp) (2002).

\bibitem{Vedral:1997}
Vedral, V., Plenio, M.~B., Rippin, M.~A., and Knight, P.~L., ``Quantifying
  entanglement,'' {\em Phys. Rev. Lett.}~{\bf 78},  2275--2279 (1997).

\bibitem{Penrose:1967}
Kronheimer, E. and Penrose, R., ``On the structure of causal spaces,'' {\em
  Proc.Cambridge Phil.Soc.}~{\bf 63},  481--501 (1967).

\bibitem{Major:2007}
Major, S., Rideout, D., and Surya, S., ``{On recovering continuum topology from
  a causal set},'' {\em J. Math. Phys.}~{\bf 48}(3),  2501 (2007).

\bibitem{Mischaikow:CH}
Kaczynski, T., Mischaikow, K., and Mrozek, M.,  [{\em Computational Homology
  (Applied Mathematical Sciences)}{\nolinebreak\hspace{0.1em}]}, Springer,
  1~ed. (Jan. 2004).

\bibitem{Zomorodian:CompTopo09}
Zomorodian, A., ``{Computational Topology},'' {\em Algorithms and Theory of
  Computation Handbook} ,  395 (2009).

\bibitem{Carlsson:2009}
Carlsson, G., ``{Topology and data},'' {\em Bull. Amer. Math. Soc.(NS)}~{\bf
  46}(2),  255--308 (2009).

\bibitem{Alexandrov:CombTopo}
Aleksandrov, P.,  [{\em Combinatorial Topology}{\nolinebreak\hspace{0.1em}]},
  vol.~1, Graylock Press, Baltimore (1956).
\newblock translation by H. Komm.

\bibitem{Zeilinger:QCA}
Gr{\"o}ssing, G. and Zeilinger, A., ``{Quantum cellular automata},'' {\em
  Complex Systems}~{\bf 2}(2) (1988).

\bibitem{Watrous:QCA}
Watrous, J., ``{On one-dimensional quantum cellular automata},'' {\em Proc. of
  the 36th Annual Symposium on Foundations of Computer Science} ,  528--537
  (1995).

\bibitem{Schumacher:QCA}
Schumacher, B. and Werner, R.~F., ``{Reversible quantum cellular automata},''
  {\em arXiv.org}~{\bf quant-ph},  5174 (2004).

\bibitem{PDC:QCA}
Perez-Delgado, C.~A. and Cheung, D., ``{Local unitary quantum cellular
  automata},'' {\em Phys. Rev. A}~{\bf 76}(3),  32320 (2007).

\bibitem{Brennen:QCA}
Brennen, G. and Williams, J., ``{Entanglement dynamics in one-dimensional
  quantum cellular automata},'' {\em Phys. Rev. A}~{\bf 68}(4) (2003).

\bibitem{Wheeler:itfrombit}
Wheeler, J., ``{How Come the Quantum?},'' {\em Annals of the New York Academy
  of Sciences}~{\bf 480}(1),  304--316 (1986).

\bibitem{Jonckheere:NetCurv}
Jonckheere, E. and Lohsoonthorn, P., ``{Geometry of network security},'' {\em
  Proc. of the 2004 American Control Conference}~{\bf 2},  976--981 (2004).

\bibitem{Hamilton:RF}
Hamilton, R., ``{Three-manifolds with positive Ricci curvature},'' {\em J.
  Differential Geom}~{\bf 17}(2),  255--306 (1982).

\end{thebibliography}
\bibliographystyle{spiebib}   %>>>> makes bibtex use spiebib.bst

\end{document}